\documentclass[9pt,twocolumn,twoside]{osajnl}

\journal{jocn} 

\setboolean{shortarticle}{false}

\usepackage[customcolors]{hf-tikz}
\usepackage{amsmath}

\newcommand{\insilex}{\emph{insilex}}
\newcommand{\NAi}{$\mathsf{NA}_\mathrm{i}$}
\newcommand*{\defeq}{\stackrel{\text{def}}{=}}
\newcommand*{\revision }{}

\title{Cross-grating phase microscopy (CGM): \emph{In silico} experiment (insilex) algorithm, noise and accuracy}

\author[1]{Baptiste Marthy}
\author[1,*]{Guillaume Baffou}
\affil{Institut Fresnel, CNRS, Aix-Marseille University, Centrale Marseille, Marseille, France}

\affil[*]{guillaume.baffou@fresnel.fr}

\begin{abstract}
Cross-grating phase microscopy (CGM) is a quantitative phase microscopy technique based on the association of a 2-dimensional diffraction grating (cross-grating) and a regular camera sensor, separated by a millimetric distance. This simple association enables the high-resolution imaging of the complex electric field amplitude of a light beam (intensity and phase) from a single image acquisition. While CGM has been used for metrology applications in cell biology and nanophotonics this last decade, there has been few studies on its basics, especially for the microscopy community. In this article, we provide a numerical algorithm that enables the \emph{in silico} {\revision  (i.e. computer-simulated)} data acquisition, to easily vary and observe the effects of all the CGM experimental parameters using computer means. In the frame on this article, we illustrate the interest of this numerical algorithm by using it to explain and quantify the effects of several important CGM parameters (grating-camera distance, pixel size, light intensity, numerical apertures, etc) on the noise, precision and trueness of CGM measurements. This work is aimed to push the limits of CGM toward advanced applications in biomicroscopy and nanophotonics. 
\end{abstract}

\setboolean{displaycopyright}{true}

\begin{document}

\maketitle

\section{Introduction}

Quantitative phase microscopy (QPM) refers to techniques capable of mapping the phase of a light beam \cite{PO57_133,OLE135_106188,NP12_578} using optical microscopy means. Many different QPM techniques implemented on optical microscopes have been developed and improved these last two decades, {\revision  e.g.}, digital holographic microscopy (DHM) \cite{AO38_6994,OL30_468}, spatial light interference microscopy (SLIM) \cite{OE19_1016,AOP13_353}, diffraction phase microscopy (DPM) \cite{AOP6_57,AO46_A52}, shack-Hartmann wavefront sensing\cite{OL42_2122,OC222_81,OE29_5193} and quadriwave lateral shearing interferometry (QLSI) \cite{JPDAP54_294002,OE17_13080}. QLSI is a QPM technique based on the association of a 2-dimensional diffraction grating (aka cross-grating) and a regular camera, separated by a millimetric distance \cite{JPDAP54_294002}. This type of cross-grating phase microscopy (CGM), introduced and patented in 2000 by Primot et al. \cite{AO32_6242,AO39_5715}, has been used for the first time on a microscope in 2009,  for bio-cell imaging and characterization \cite{OE17_13080,JBO20_126009}, and more recently in nanophotonics for the characterization of the optical properties of nanoparticles \cite{JOSAA36_478,O7_243,PRB86_165417}, 2D-materials \cite{ACSP4_3130} and metasurfaces \cite{ACSP8_603}, and as a temperature microscopy technique in nanoplasmonics \cite{ACSNano6_2452,ACSNano7_6478,N6_8984,S14_1801910,JPCC125_21533}. Despite the gain in popularity of CGM, its basics remain poorly investigated in the field of optical microscopy.

In this article, we introduce an image processing algorithm that enables the simulation of experimental CGM images, taking into account the camera shot noise and light beam propagation between the cross-grating and the camera chip. Numerical results are compared with experimental measurements using a home-made, tunable CGM set up to validate the algorithm. In a large part of the article, using this algorithm, we discuss the influence of parameters such as light intensity, grating-camera distance, relay lens magnification, numerical aperture (NA) of the illumination, on the image noise level, measurement precision and trueness. {\revision  The algorithm is attached in Suppl. Info. and accessible on a public repository \cite{GitHub_CGMinSilico}.}

\section{Cross-grating phase microscopy (CGM)}
\subsection{Basic principle of CGM}
\begin{figure*}
	\centering
		\includegraphics[scale=1]{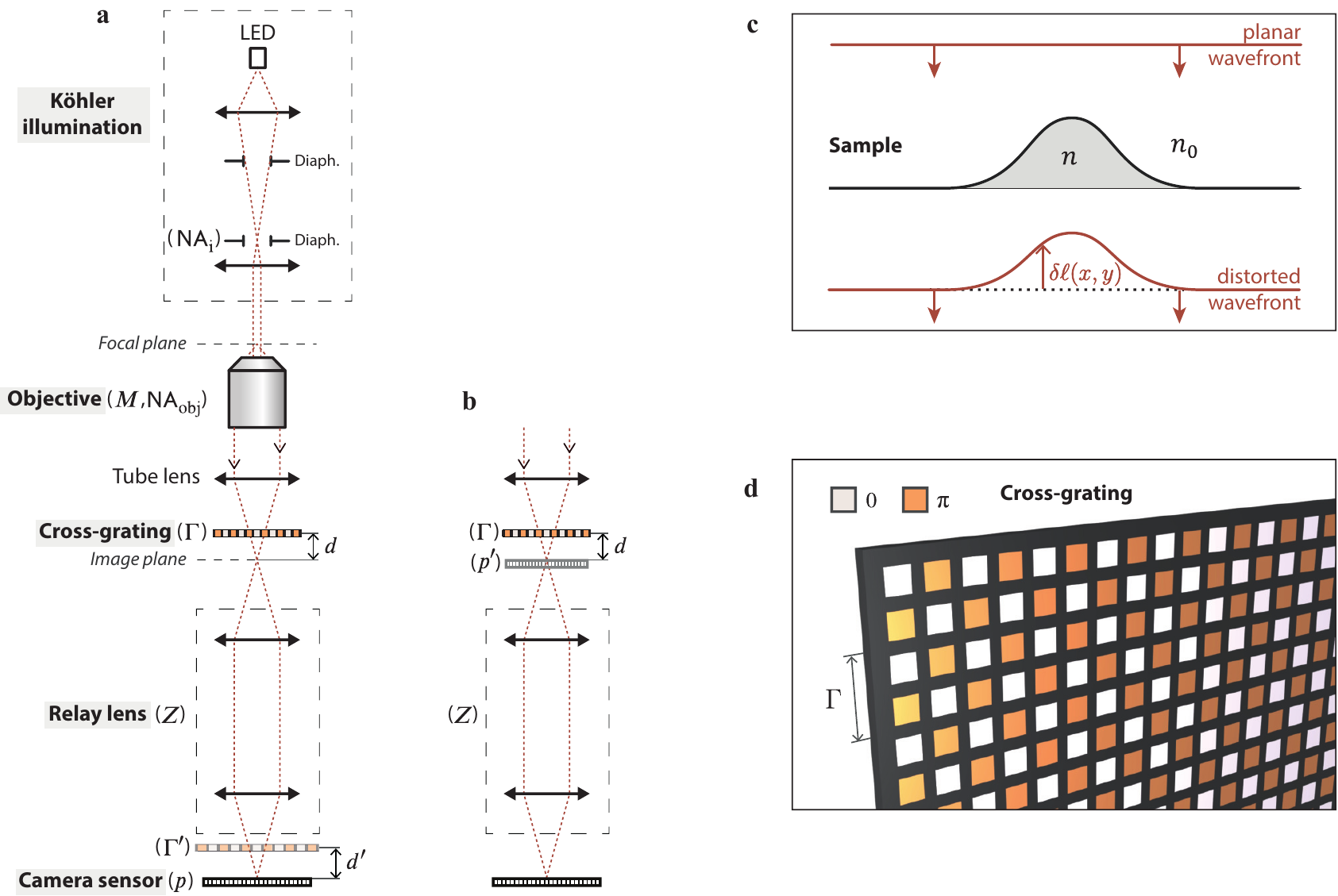}
	\caption{{\bfseries Working principle of cross-grating microscopy.} (a) Schematic of the experimental set up {\revision  where the grating is considered to be imaged by the relay lens, at the vicinity of the camera sensor (grating-image description). (b) Representation of the imaging part of the setup where the relay lens is considered as an imager of the camera sensor instead of the grating (sensor-image description).} (c) Optical wavefront distortion due to the presence of a transparent object, defining the optical path difference $\delta\ell$. (d) Representation of a QLSI cross-grating, characterized by a $0-\pi$ checkerboard pattern.}
	\label{setup}
\end{figure*}

Cross-grating phase microscopies (CGMs) use a 2D-dimensional diffraction grating (aka a cross grating \cite{book_BornWolf}) positioned at a millimetric distance from the chip of a regular camera \cite{JPDAP54_294002}. Common cameras do not offer the possibility to place an object so close to the chip, due to presence of a sealed chamber. While some built-in commercial CGM systems exist, home-made systems rather involve a relay lens system that re-image the cross-grating at the desired distance from the camera chip (Figure \ref{setup}a).
The cross-grating creates a so-called interferogram image that is processed in real time to retrieve both the intensity and the phase of the incoming light beam. Unlike most QPMs, which directly measure and map the phase of light $\phi$, CGM primarily measures the wavefront profile $W$ of a light beam, or rather its gradients over the two directions of the focal plane (that are subsequently integrated). The interferogram consists of a dense array of bright spots, and the working principle resembles that of a Shack-Hartmann sensor \cite{OL42_2122,OC222_81,OE29_5193}, although with a much higher spatial resolution. Then, the wavefront profile $W$ can be converted into the phase profile, if need be, using the relation
\begin{equation}
\phi(x,y)=\frac{2\pi}{\lambda_0}W\left(x,y\right),
\label{eq:phi}
\end{equation}
where $\lambda_0$ is the illumination wavelength. Thus, CGM is primarily a wavefront sensing technique. Its consideration as a phase microscopy technique is rather a means to make it more popular in the field of bio-microscopy. {\revision  Note that Eq. \eqref{eq:phi} assumes a monochromatic light, or at least a wavelength range $\lambda_0\pm\Delta\lambda/2$ over which the imaged object is not too dispersive.}

CGM was not originally developed to be plugged onto an optical microscope. The original purpose of CGM was rather to characterize the quality of laser beams \cite{OL23_621}. The idea to plug a CGM camera into a microscope to use it as a QPM for bioimaging was introduced by Bon et al. in 2009 \cite{OE17_13080}. In this case, the wavefront profile results from a distortion due to the presence of a refractive object (a bio-cell) at the sample plane of the microscope, and is called the optical path difference (OPD) $\delta\ell=W$  (Figure \ref{setup}b) and reads
\begin{equation}
\delta\ell(x,y)=\left(n-n_0\right)h\left(x,y\right)
\end{equation}
where $h$ is the thickness profile of the object, $n$ the refractive index of the object and $n_0$ the refractive index of the environment.

{\revision In practice, to compensate for any imperfection of the incoming planar wavefront, a reference interferogram is recorded first, without the object of interest in the field of view, from which a reference OPD is calculated and subtracted from all subsequent images of interest.}

Different types of cross-gratings have been used in CGM, with 3-fold or 4-fold symmetries \cite{JOSAA12_2679}, with different designs of the unit cell \cite{OC451_86,OE26_26872,JO_Li_2022,OL40_2245}, or even with a thin (non-periodic) diffuser \cite{OL42_5117} or a binary random mask \cite{SR9_13795}. In the context of optical microscopy, the main instance of CGM that has been used so far is quadriwave lateral shearing interferometry (QLSI) \cite{JPDAP54_294002,OE17_13080,JOSAA12_2679}. QLSI cross-gratings feature a 4-fold symmetry consisting of horizontal and vertical opaque lines defining transparent squares imprinting $0$ and $\pi$ phase shifts on the incoming light according to a checkerboard pattern  (Figure \ref{setup}c).

\subsection{CGM experimental set up}
Although this article is mainly aiming at introducing a numerical algorithm, simulations will be compared with experimental measurements. To conduct these experiments, we used a home-made CGM set up, composed of a QLSI cross-grating, with a period of $\Gamma=39$ \textmu m and a Sona camera from Andor {\revision ($2048\times2048$ dexels, dexel size $p=6.5$ \textmu m)}. The grating was re-imaged using a relay lens (VZM 300, zoom 
$0.75\times$ - $3\times$, Edmund optics, ref. $\sharp$39-708)  and the grating-camera distance $d$ was controlled using a stepper motor actuator (Thorlabs {\sf LNR25ZFS/M}, {\sf KST101}) (Fig. \ref{setup}a). The microscope was also home-made, composed of a $60\times$ objective lens (Olympus, {\sf LUCPLFLN60X}) and a $180$-mm tube lens (Thorlabs, {\sf TTL200-A}). The sample was illuminated using a mounted LED at $625\pm25$ nm (Thorlabs, {\sf M625L3}), associated with a K\"ohler configuration.

\subsection{Experimental parameters in CGM}

A CGM setup as depicted by Fig. \ref{setup}a can be customized by varying several parameters, listed in Table \ref{tbl1}, related to the features of the microscope, the grating and the camera. While commercial CGM systems are usually fixed, home-made CGM systems offer the possibility to vary and optimize the geometrical parameters $\mathcal{U}$, $\Gamma$, $d$, $\beta$, $Z$ {\revision (defined in Table \ref{tbl1} and later in the text)}, depending on the application and the sample. Optimizing experimentally the 13 parameters listed in Table \ref{tbl1} may be cumbersome, hence the interest of conducting \emph{in silico} experiments. Let us review and comment all the parameters of Table 1 one by one.

\begin{table*}
\footnotesize
\centering
\caption{Definitions of the physical parameters involved in CGM, along with their particular values used in all the figures of this article.}\label{tbl1}
\begin{tabular*}{5.2in}{llllllll}
\hline

Param. & Definition & Unit & Fig. \ref{algoStructure}\&\ref{compOPD}& Fig. \ref{noiseImages}& Fig. \ref{sigma_distance}& Fig. \ref{sigma_NA}& Fig. \ref{accuracy}\\
\hline
$\lambda_0$ &	Light wavelength &	nm	& $530$ & $625$& $625$& $625$& $650$\\
$e_0$ &	Etching depth of the cross-grating &	nm	& $530$ & $625$& $625$& $625$& $650$\\
$w$ &	Maximum counts on the interferogram& &	$45000$ & $40150$& $40000$& $40000$&  $200$\\
& \emph{(full well capacity of the camera)} &&\\
$N_\mathrm{im}$ &	Number of averaged interferograms& & $5$ & $30$& $25$& $25$& $1$\\
$p$ &	Pixel size of the camera (dexel size) &	\textmu m &	$6.5$ &	$6.5$&	$6.5$&	$6.5$&	$6.5$\\
$\mathcal{U}$ &	Grating unit cell (grexel) &&QLSI&QLSI&QLSI&QLSI&QLSI\\
$\Gamma$ & Grating period (grexel size) &	\textmu m	& $39$	& $39$& $39, 52, 65$& $39,52$& $52$\\
$Z$ &	Zoom of the relay lens & & $1$ & $1$ & $1$ & $1$ & $1$\\
$d$ &	Grating-camera distance	 & mm	& $1$	& $1.03$ & $0.3-3$& $0.5-3$& $0.2-2.2$\\
$\beta$ &	Grating tilt angle	& deg	& $37^\circ$& $37^\circ$& $37^\circ$& $37^\circ$& $37^\circ$\\
\NAi\ &	Numerical aperture of the illumination	& & $0$& $0$& $0$& $0-0.9$& $0$\\
$N_x, N_y$ &	Size of the OPD image	& px & $240$ & $720$& $600$& $600$& $600$\\
$M$ &	Magnification of the microscope	&  & & & & $100$& $40$\\
\hline
\end{tabular*}
\end{table*}

\begin{itemize}
\item$\lambda_0$. CGMs are achromatic. The knowledge of the illumination wavelength $\lambda_0$ is not necessary to compute the wavefront profile from the interferogram, and varying the wavelength is not supposed to change the OPD profile retrieved from the interferogram. However, the phase shifts ($0$ and $\pi$ in QLSI) are imprinted on the grating by etching the substrate. Thus, the phase shifts are supposed to depend on the wavelength. However, if $\lambda_0$ deviates from the wavelength the grating has been made for, it is not supposed to lead biased measurement, but only poorer signal to noise ratio. Also, CGM does not require the use of coherent light sources (laser), unlike other QPM techniques. On the contrary, it is even recommended to use temporally incoherent, broad band, light sources rather than a laser light to avoid the appearance of fringes on the intensity and phase images. {\revision  Numerically, however, the use of a coherent light source description does not cause problem.}
{\revision  \item $e_0$. The phase pattern of a cross-grating used in CGM is made by local etching of the substrate. For a QLSI pattern, the $\pi$ phase shift are obtained by etching the substrate over a distance $e_0$ such that
\begin{equation}
\pi=\frac{2\pi}{\lambda_0}(n_0-1)e_0.
\end{equation}
In the manuscript, we consider that the substrate refractive index is $1.5$, so that $\lambda_0=e_0$ to obtain a $\pi$ phase shift.}
\item $w$ is the value of the brightest pixels of the interferogram (in photo-electrons). This value has to be adjusted just below the full well capacity of the camera chip to optimize the signal to noise ratio. In this article, we consider $w$ to be the full well capacity. Typical scientific cameras feature a full well capacity ranging from $10000$ to $50000$, encoded in 16-bit.
\item $N_\mathrm{im}$ is the number of averaged interferograms, or equivalently of OPD images (averaging one or the other has an equivalent effect on the noise amplitude).
\item $p$ is the lateral size of the camera dexel. A dexel means 'detector element' \cite{book_Bushberg}, just like a pixel means a 'picture element'. We opt for this appellation instead of a camera 'pixel' to avoid confusion with the pixel of an image.
\item $\mathcal{U}$ is the grating unit cell. Following the same logic, we shall call it a grexel (grating element). In this article, we focus on the cross-grating used in QLSI, with a grexel characterized by a checkerboard pattern of $0$ and $\pi$ phase shifts. The grexel pattern of QLSI remained mostly unchanged for 20 years, except in few articles \cite{OC451_86,OE26_26872,JO_Li_2022,OL40_2245,SR9_13795}.
\item $\Gamma$ is the grexel lateral size, i.e., the grating period. It is not a parameter that can be easily and continuously modified experimentally, highlighting the interest of conducting \emph{in silico} experiments, prior to the design and fabrication of a CGM grating.
\item $d$ is the distance between the grating and the image plane. $d$ is an important parameter that affects both the precision and trueness of the measurements, as explained hereinafter. It usually lies in the millimetric range.
\item $Z$ is the magnification of the relay lens. {\revision   Its role can be understood in two different ways, following the two descriptions of Figs. \ref{setup}a and \ref{setup}b. First (Fig. \ref{setup}a), the relay lens can be seen as a means to image the cross-grating at a distance \cite{JOSAA36_478}
\begin{equation}
d'=Z^2d
\end{equation} 
from the camera sensor and expand it by a factor of $Z$, leading to an effective grexel size of
\begin{equation}
\Gamma'=Z\Gamma.
\end{equation}
This is the common vision.\\
Reciprocally, the relay lens can be seen as a means to image the \emph{camera} at a distance $d$ from the actual cross-grating, and scale it by a factor $1/Z$. This second vision, less intuitive, is actually much simpler: there is only one distance to consider, $d$, the one that is experimentally actuated (no need to worry about another distance and about a factor of $Z^2$), and it simplifies the algorithm as the role of the relay only amounts to only scaling the dexel size $p$ of the camera by a factor of $1/Z$:
\begin{equation}
p'=p/Z.
\end{equation}
} An important feature of a CGM system, as explained later on, is the $\zeta$ (zeta) factor, that is the ratio between the effective grexel size $\Gamma'=Z\Gamma$ and twice the camera dexel size $p$:
\begin{equation}
\zeta=\frac{Z\Gamma}{2p}.
\end{equation}
{\revision  Note that this ratio can also be seen as the actual grexel size $\Gamma$ divided by twice the effective dexel size $p/Z$, stressing the fact that the grating-image and sensor-image descriptions are equivalent (Figs. \ref{setup}a and Figs. \ref{setup}b).} Common CGM cameras usually use $\Gamma'=6p$ or $8p$, i.e., $\zeta=3$ or $4$. The $\zeta$-factor must remain greater than $2.73$ (Nyquist criterion) to enable proper imaging of the grexels, i.e., proper sampling of the fringes of the interferogram \cite{OE17_13080}.
\item $\beta$ is the rotation of the grating around the optical axis. In practice, although CGM is supposed to normally work with $\beta=0$, the cross-grating is usually tilted by an angle $\beta\neq0$ around the optical axis to avoid Moir\'e effects and OPD reconstruction issues. For this reason, the algorithm we propose offers this option.
\item $\mathsf{NA}_\mathrm{i}$ is the NA of the illumination. It is an important parameter that can affect the signal to noise ratio of the reconstructed OPD image \cite{OE17_13080}.
\item $(N_x,N_y)$ are the dimensions, in pixel, of the OPD image. These dimensions are supposed to equal the ones of the interferogram, except for some algorithms that reduce the size of the OPD image to $(N_x/\zeta,N_y/\zeta)$, for the good of the speed for live imaging, but at the expense of the image quality. In this article, we always process the OPD images so that they feature the same number of pixels as the interferogram, according to the algorithm we recently detailed in Ref. \cite{JPDAP54_294002}.
\item $M$ is the magnification of the microscope. It only matters when dealing with the illumination aperture or accuracy issues (see sections \ref{sect:NA} and \ref{sect:trueness}).
\end{itemize}

\section{Numerical procedure}
\begin{figure*}
	\centering
		\includegraphics[scale=1]{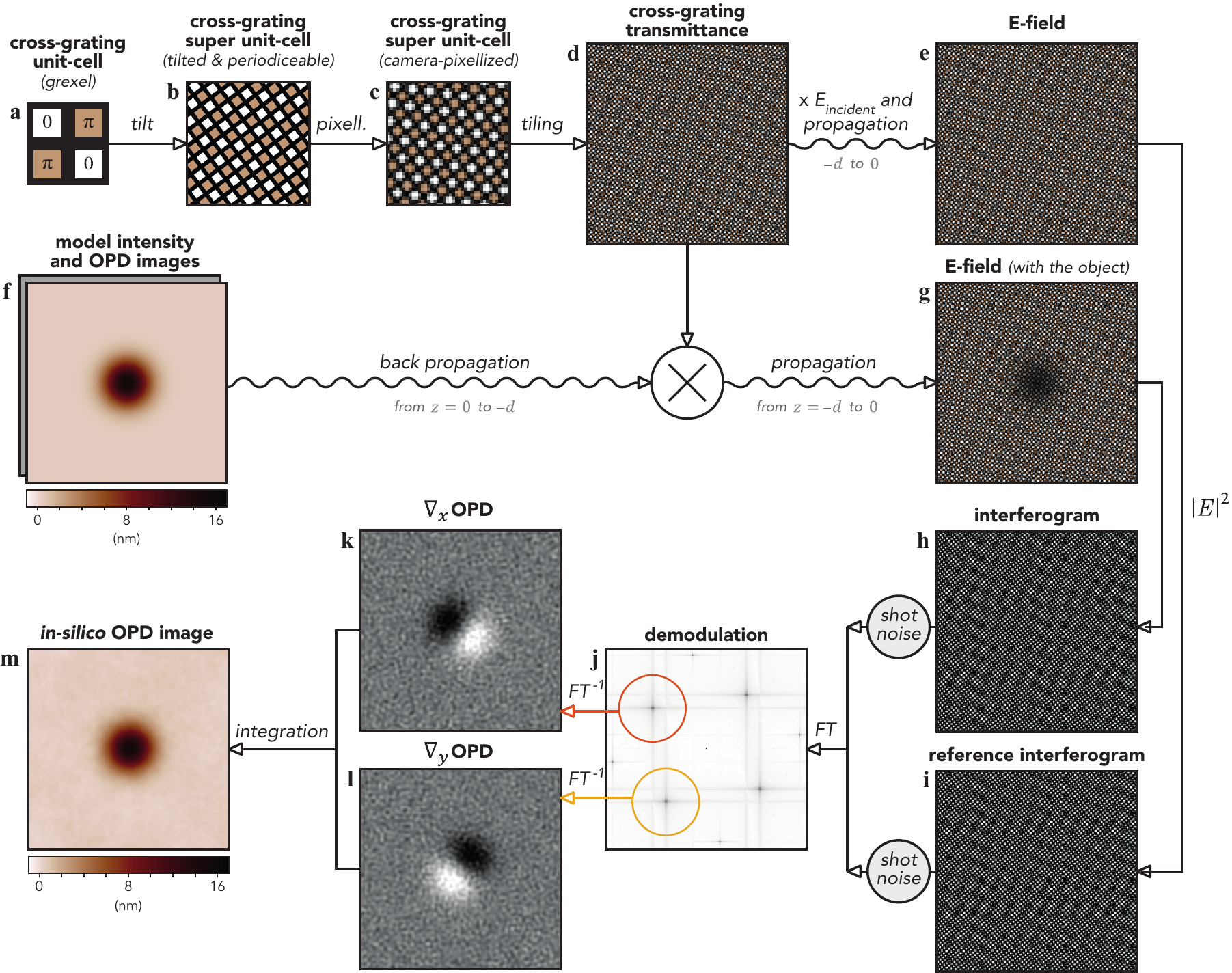}
	\caption{{\bfseries Schematic of the algorithm for \emph{in silico} experiments (\insilex\  algorithm).} (a) Unit cell of the cross grating. Black lines are zero-transmission areas. (b) Tilt of the unit-cell by $\beta=\cos^{-1}(3/5)$ and enlargement by a factor of 5 to obtain a periodiceable super unit-cell. (c) Pixelization of the super unit-cell according to the dexel size of the camera. (d) Tiling of the super unit-cell, up to the size of the camera chip. (e) Numerical propagation of the electric field from the grating to the camera chip. (f) Manually designed (model) intensity and OPD $\delta\ell$ images (here a Gaussian OPD profile). (g) Multiplication of the $\exp{(i2\pi\delta\ell/\lambda_0)}$ image by the cross grating transmittance. (h) Interferogram and (i) reference interferogram followed by shot noise addition and 2D Fourier transform. (j) Demodulation of the interferograms. (k,l) Inverse Fourier transform and integration to obtain the \emph{in silico} measured OPD (m), to be compared with (f).}
\label{algoStructure}
\end{figure*}

{\bfseries Figure \ref{algoStructure}} depicts the algorithm we developed and introduce in this article to simulate experimental CGM images. We coin it the \insilex\  (\emph{in silico} experiment) algorithm. {\revision It is composed of two parts, the traditional part (steps h to m) that processes the interferogram to retrieve the OPD \cite{JPDAP54_294002}, and a novel part (steps a to g + noise addition between steps i and j) aimed to simulate experimental images.} Here is a detailed description of this algorithm.\\

\noindent{\bfseries Figure \ref{algoStructure}a}: First, the grexel pattern is designed. We focus in this article on a QLSI grexel, characterized by a $0$-$\pi$ checkerboard pattern. The aspect ratio of the dark lines width and the grexel size is 1/6, a value aimed at optimizing the emission of light on the first orders of diffraction \cite{AO39_5715}.\\
{\bfseries Figure \ref{algoStructure}b}: The grexel should be tiled to form the full grating. Prior to tiling, one apply a tilt $\beta$ of the grexel. A tilt by an arbitrary angle would produce discontinuities of the periodicity at the junction of the grexel tiles, upon tiling. To avoid this issue, we chose to tilt the grexel by a particular angle of $\beta=\cos^{-1}(3/5)$. This angle value ensures continuity if the unit-cell is made exactly 5 times bigger, leading to what we call the super unit-cell. This magic angle yielding continuity upon tiling comes from the integer equality $3^2+4^2=5^2$.\\
{\bfseries Figure \ref{algoStructure}c}: The grating super unit-cell is then resampled so that the pixel density matches the dexel density of the camera chip.\\
{\bfseries Figure \ref{algoStructure}d}: Super unit-cells are tiled to get the complex transmittance $T_\mathrm{g}$ image of the full grating, of the size of the camera chip.\\
{\bfseries Figure \ref{algoStructure}e}: The transmittance image $T_\mathrm{g}$ is multiplied by the electric field amplitude of the uniform incoming light beam, possibly tilted by a deviation angle $\psi$ from the optical axis, and then propagated over a distance $d$ to get the reference $E$-field on the image plane. Considering a set of various illumination angles $\psi$ enables the modelling of $\mathsf{NA}_\mathrm{i}\neq0$.\\
{\bfseries Figure \ref{algoStructure}f,g}: Meanwhile, intensity $I_0$ and phase $\phi_0$ profiles are designed, corresponding to a desired object to be imaged. We call them the \emph{model} intensity and \emph{model} phase images. Then, the scalar field $\sqrt{I_0}\exp(i\phi_0)$ is backward-propagated from the image plane to the grating, multiplied by the grating transmittance $T_\mathrm{g}$, and forward-propagated to get the $E$-field at the image plane. Back and forth propagations are simulated using a standard Fourier-transform algorithm (see Matlab code \texttt{improp.m} in Suppl. Info.).\\
{\bfseries Figure \ref{algoStructure}h,i}: The two interferograms are calculated from the two $E$-field maps, with and without the imaged object. Shot noise is added to the image. The amplitude of the shot noise is directly related to the number of counts on each dexel. In the \insilex\ code, it is created using the \texttt{poissrnd} function of Matlab.\\
{\bfseries Figure \ref{algoStructure}j,k}: Then, the home-made standard algorithm that we normally use to postprocess experimental interferograms is here applied to the simulated interferograms to retrieve the intensity $I$ and phase $\phi$ images, to be compared with the model $I_0$ and phase $\phi_0$ images. This algorithm consists in a demodulation of the image in the Fourier space to retrieve two wavefront gradients along orthogonal directions, which are then integrated to retrieve the wavefront profile (see Ref. \cite{JPDAP54_294002} for details and the \texttt{CGMprocess.m} Matlab code in Suppl. Info). Note that the postprocessing algorithm can also retrieve the intensity map from the interferogram by cropping the central spot in the Fourier space {\revision \cite{JPDAP54_294002}} (not shown in Figure \ref{algoStructure} for the sake of simplicity).

\begin{figure}[!btp]
	\centering
		\includegraphics[scale=1]{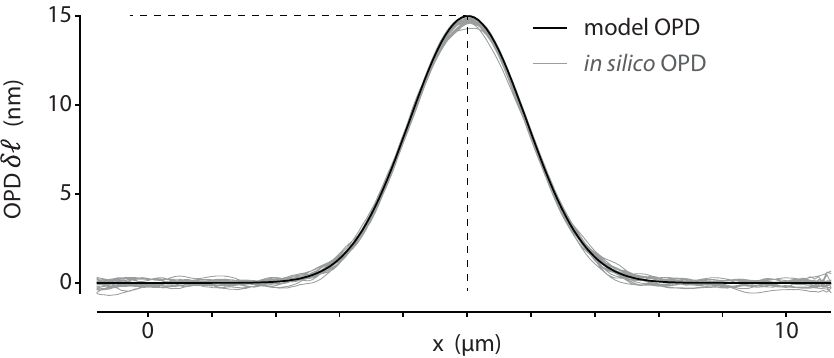}
	\caption{Numerical simulation of the model OPD profile, and 17 experimental OPD profiles produced \emph{in silico}.
}
\label{compOPD}
\end{figure}

A Matlab package reproducing this algorithm is provided in Suppl. Info {\revision  and on a public repository \cite{GitHub_CGMinSilico}}. As an initial test of thec\insilex\  algorithm, Fig. \ref{compOPD} plots the crosscut of the model OPD shown in Figure \ref{algoStructure}f (a simple Gaussian profile, 15 nm in amplitude), in comparison with a series of \emph{in silico} calculated OPDs, where a proper agreement is observed. Experimentally, such a good agreement is not always ensured. Further dispersion of the measurements (lack of precision) and measurement bias (trueness issue) can be encountered if the set of experimental parameters listed in Table 1 is not properly adjusted. These experimental limitations can be rendered by the \insilex\ algorithm. Next sections discuss these limitations to illustrate the interest of \emph{in silico} experiments, and to eventually better define the best working area of CGMs, and QLSI in particular.

{\revision Using the Matlab algorithm, each image requires around 1 s to be processed (1.25 second for a $600\times600$ px image) with a standard desktop computer. The computation time is proportional to the number of pixels of the image ($t\sim N_xN_y$). The shot noise generation, via the use of the Matlab function \texttt{poissrnd}, is quite time-consuming, responsible for 2/3 of the total computation time.}

\section{Image noise and precision}
Precision, trueness and accuracy are important to determine when conducting experimental measurements, and quantitative phase microscopy is no exception \cite{JBO20_126009}.

{\bfseries Precision} refers to the standard deviation of an ensemble measurements performed in the same experimental conditions, for instance coming from the noise on the image. {\bfseries Trueness} is the deviation of the measurements from the true value, also called a bias. {\bfseries Accuracy}, sometimes mixed with the trueness, normally encompasses both precision and trueness. This section focuses on the estimation of the precision, while the next one focuses on the trueness.

\begin{figure*}
	\centering
		\includegraphics[scale=1]{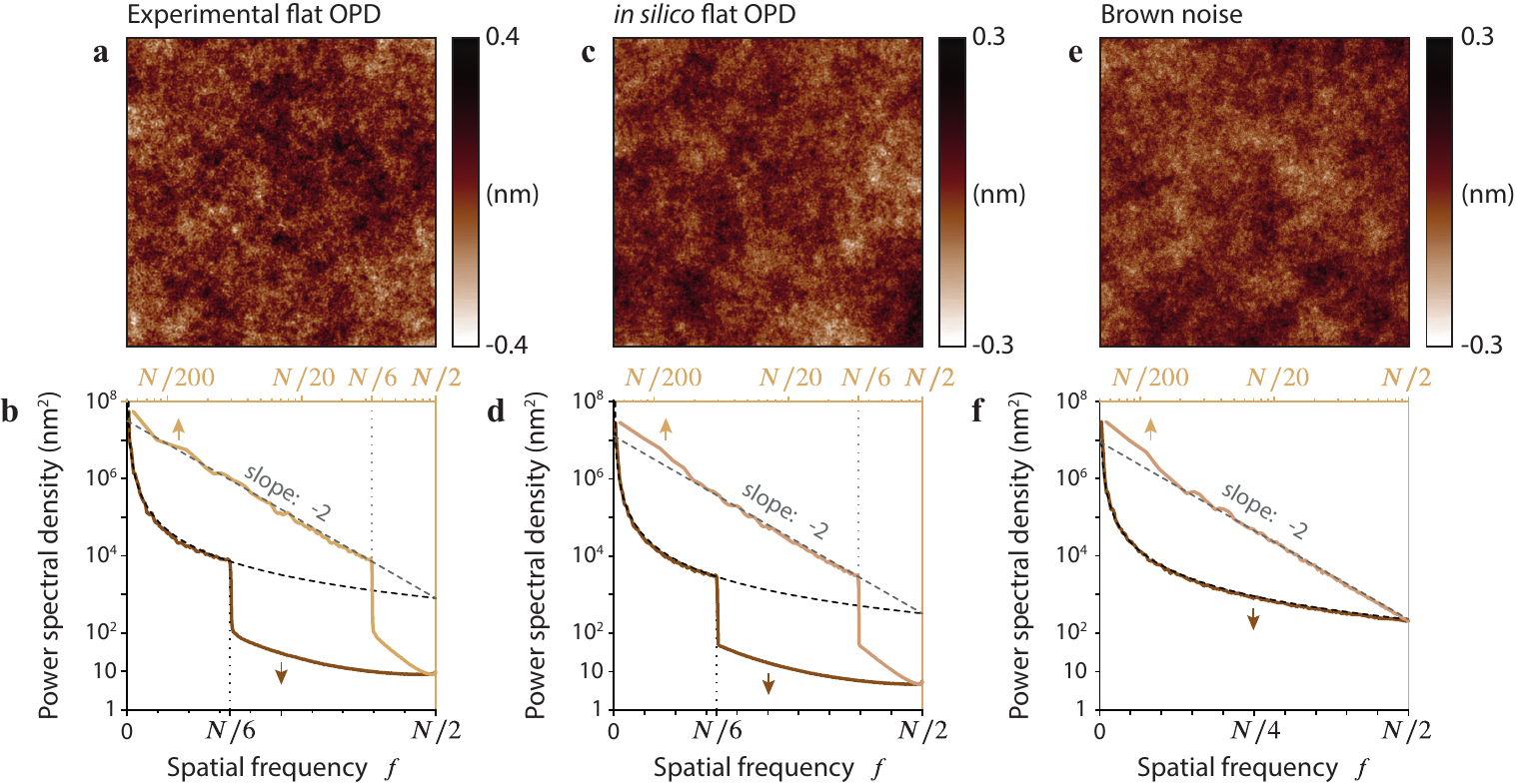}
	\caption{{\bfseries Characterization of the image noise.} (a) Experimental uniform OPD. (b) Spatial frequency spectrum of image (a).  (c,d) Same as (a,b) for an \emph{in silico} uniform OPD image in the same conditions. (e) Theoretical image featuring Brown noise. (f) Spatial frequency spectrum of the image (e). }
\label{noiseImages}
\end{figure*}

\subsection{White noise or Brown noise?}
Figure \ref{noiseImages} displays experimental and numerical flat OPD images (no object is imaged), as a means to highlight the image noise. Figure \ref{noiseImages}a shows an experimental OPD image (see Table \ref{tbl1} for details on the experimental conditions), along with its power spectral density (Fig. \ref{noiseImages}b). The noise in CGM mainly comes from the shot noise on the camera (aka photon noise). Although a shot noise is a white noise (no spectral dependence on the spatial frequencies), OPD images in CGM do not feature a white noise, but rather a noise characterized by a $1/f^2$ power spectral density (Fig. \ref{noiseImages}b) where $f$ represents the spatial frequencies of the image. A $1/f^2$ noise is usually called a Brown or Brownian noise. This particular noise arises from the integration step (Fig. \ref{algoStructure}k,l,m), not from the demodulation (Fig. \ref{algoStructure}j), as OPD gradients (Figure \ref{algoStructure}k,l) still feature a white noise. A Brown noise is indeed obtained by integrating a white noise.

Note the cutoff frequency at $f=N/6$, where $N\;\defeq\;N_x=N_y$, in Fig. \ref{noiseImages}b. This cutoff comes from the demodulation step (Fig. \ref{algoStructure}j) that consists in cropping the Fourier space by a disc of diameter $N/\zeta$.

Figure \ref{noiseImages}c displays an OPD image calculated using the \insilex\ algorithm for the exact same parameters as Fig. \ref{noiseImages}a. This image along with its power spectral density plot (Fig. \ref{noiseImages}d) properly reproduce the experimental Brown nature of the noise (see Fig. \ref{noiseImages}a,b).  As an illustration, a theoretically generated Brown noise is also shown in Figure \ref{noiseImages}e, which visually and spectrally renders the same characteristics, except that no cutoff frequency has been applied.

The noise standard deviation is slightly better (i.e. lower) in the \insilex\ image, $\sigma_0=67$ pm, compared with the experimental image, $\sigma=97$ pm. In CGM, other sources of noise can be a setup misalignment, aberrations, sensor non-linearity, or a non-zero illumination NA (for the latter, see section \ref{sect:NA}).

\subsection{Noise estimation in CGM}
The \insilex\ algorithm enables the determination of the fundamental, minimum noise standard deviation $\sigma_0$ that can be achieved in a CGM experiment as a function of all the experimental parameters. Measuring $\sigma>\sigma_0$ means that the setup can be further optimized. We conducted a large amount of \insilex\ calculations, varying all the parameters, to understand all the dependencies of the parameters and derive a semi-empirical expression for $\sigma_0$:
\hfsetfillcolor{gray!8}
\hfsetbordercolor{black!50}
\begin{equation}\tikzmarkin{b}(0.28,-0.6)(-0.20,0.85)
\sigma_0=\frac{1}{8\sqrt{2}}\frac{p \Gamma}{Zd}\sqrt{\left(\frac{1}{N_\mathrm{im}}+\frac{1}{N_\mathrm{im}^0}\right)\frac{\log(N_xN_y)}{w}}\label{eq:sigma}
\tikzmarkend{b}
\end{equation}
where $N_{\mathrm{im}}$ is the number of averaged interferograms, and $N_{\mathrm{im}}^0$ is the number of averaged reference interferograms (equation coded in \texttt{sigma0.m}, in Suppl. Info.). Eq. \eqref{eq:sigma} considers a zero illumination NA ($\mathsf{NA}_\mathrm{i}=0$). A refined expression of $\sigma_0$ including $\mathsf{NA}_\mathrm{i}$ is given later (see Eq. \eqref{eq:sigmaNA}). Other types of camera noise, like read noise or thermal noise could also contribute. However, CGMs working conditions normally involve the full well capacity of the sensor, and short exposure times, so these other types of noise are supposed to be negligible.  {\revision  Regarding the camera parameters ($p$, $N_i$, $w$), noise amplitude varies as $\sqrt{\log(N_i)}$. This dependency is expected when integrating a Brown noise. It comes from the increase in noise amplitude at lower spatial frequencies, i.e., larger image sizes. However, it does not mean that camera with few pixels must be preferred. First, this increase in the noise level with $N$ is extremely slow and then, on the contrary, the larger the field of view, the better. What matters is rather the size, in pixels, of the imaged object.  This equation also assumes that the full well capacity $w$ is used. When imaging dark objects (not fully transparent), the intensity on the camera sensor can locally decrease, artificially decreasing $w$ and locally decreasing the signal to noise ratio.}

\begin{figure*}
	\centering
		\includegraphics[scale=1]{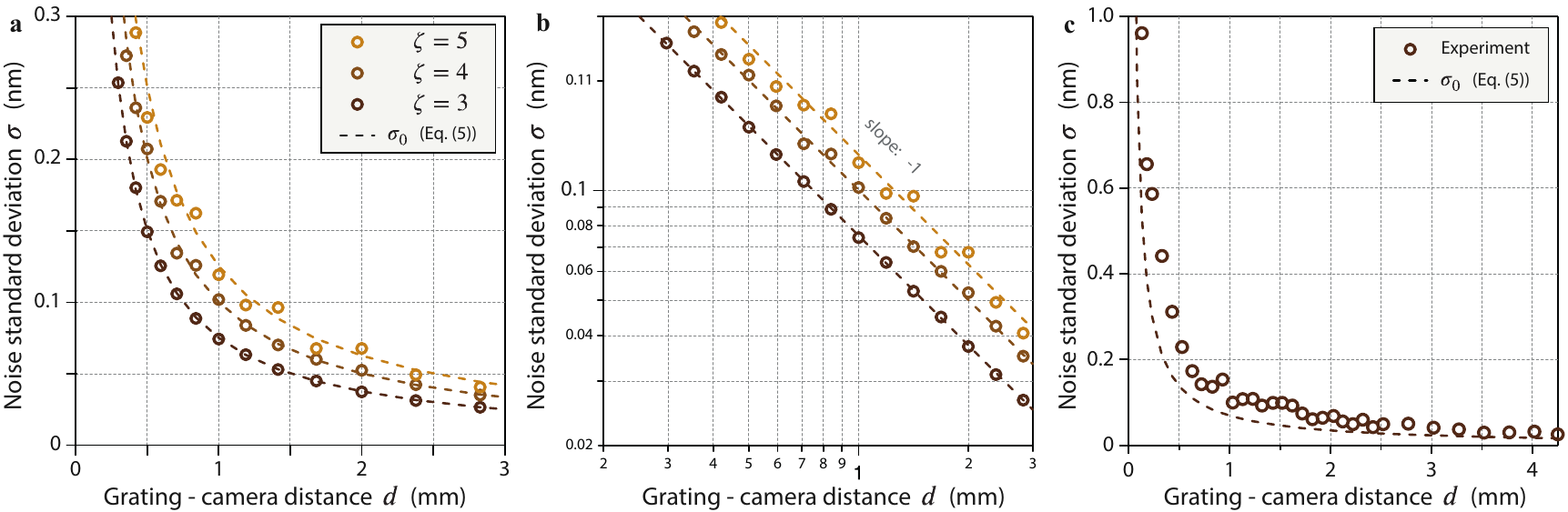}
	\caption{{\bfseries Effect of the grating-camera distance on the image noise.} (a) Noise standard deviation as a function of the grating-camera distance $d$ for different values of the $\zeta$ parameter. Dashed lines represent $\sigma$ values calculated using Eq. \eqref{eq:sigma}. (b) Same as (a) in logarithmic scale. (c) Experimental measurements of $\sigma$ compared with the minimum noise standard deviation given by Eq. \eqref{eq:sigma}.}
\label{sigma_distance}
\end{figure*}

Figure \ref{sigma_distance}a,b plots \insilex\ simulations of $\sigma$ compared with $\sigma_0$ values given by Eq. \eqref{eq:sigma}, for $\zeta=3,4,5$. A very good agreement is observed, supporting the validity of Eq. \eqref{eq:sigma}. Figure \ref{sigma_distance}c plots comparisons between $\sigma_0$ and experimental data, in the exact same conditions of parameters listed in Table \ref{tbl1}. Experimental values are very close albeit slightly higher than the fundamental limit $\sigma_0$. As mentioned above, a higher noise level can have different origins. In our case, we suppose it can come from aberrations of the optics.

One can also derive a simpler expression of $\sigma_0$, involving $\zeta$ and also considering the normal case where $N_\mathrm{im}=N_\mathrm{im}^0$:
\begin{equation}
\sigma_0=\frac{p^2 \zeta}{4Z^2d}\sqrt{\frac{\log(N_xN_y)}{N_\mathrm{im}w}}\label{eq:sigma2}
\end{equation}

It is common to read estimations of noise and precision in the literature of QPM techniques as if they were universal. However, Eq. \eqref{eq:sigma2} shows that the noise amplitude in CGM is not universal, and depends on experimental conditions and setups. {\revision  Nevetheless,  let us try to answer the question "What is the noise amplitude in CGM" with a typical value rather than an equation involving many parameters.} Noise can be quantified in several manners. First, it can be calculated for a particular image and quantified by its standard deviation, as what we have done above. However, one may want to assign a noise level not to an image, but to a technique or a particular set up, as a means to compare different techniques with each other for instance. For this purpose, caution has to be used because noise levels obviously vary for a given setup, from one set of experimental conditions to another: in particular, noise standard deviation varies with light intensity $\mathcal{I}$ and exposure time as $\sqrt{\mathcal{I}t}$. For this reason, as a means to get a more universal noise characterization taking into account the exposure time $t$ dependence, noise amplitude is often given in nm/$\sqrt{\mathrm{Hz}}$ units (for a signal in nm). However, this unit takes into account $\sqrt{t}$ but not $\sqrt{\mathcal{I}}$. The quantity $\mathcal{I}\times t$ is nothing more than the density of light energy collected by the camera. This quantity is proportional to the number of photons collected per dexel. For this reason, as a more universal figure of merit, we propose here to define the noise of CGM as the noise standard deviation of a full-frame image when the camera sensor collects 40000 photons in the brightest pixels of the interferogram. This value roughly corresponds to the full well capacity of common scientific cameras. Thus, this definition gives an idea of the noise amplitude on OPD images arising from a single interferogram acquisition. Equation \eqref{eq:sigma2} leads an estimation of this figure of merit, that is $\sigma_0=0.4$ nm/frame. To compare with other values from the literature sometimes given in nm/$\sqrt{\mathrm{Hz}}$Hz, one can consider a frame rate of $25$ Hz to derive a noise level of $\sigma_0=80$ pm/$\sqrt{\mathrm{Hz}}$, that is a fraction of the size of a hydrogen atom (120 pm).\\

{\revision  In 2013 \cite{OE21_17340}, the group of Primot demonstrated that the noise amplitude on the OPD image obtained in CGM could be estimated in a more fundamental way than using the image standard deviation. The demonstration was conducted in the frame of OPD images obtained in the X-ray spectral range, but it could be transposed in the visible range. Estimating the noise amplitude by calculating the standard deviation of the image, like what we do in this article, is appropriate when dealing with flat OPD images. In practice, flat areas may not exist on the imaged sample, and it would be meaningless to crop small areas of the field of view to calculate $\sigma_0$, because $\sigma_0$ depends on the image size $N_x\times N_y$ (see Eq. \eqref{eq:sigma}). Primot et al. have shown that the noise in the reconstructed OPD image can be directly estimated during the interferogram analysis (from Figs. \ref{algoStructure}k,l). One can define the phase derivatives closure map (PDCM) of a phase image as:
\begin{equation}
C(x,y)=\partial_x(\partial_y\Phi)-\partial_y(\partial_x\Phi)
\label{eq:PDCM}
\end{equation}
If $\Phi$ has continuous second partial derivatives at the point $(x,y)$, then $C(x,y)=0$ (Schwartz's theorem). In practice, this condition does not hold true due to image noise, and the noise level of the image is contained in $C$. Interestingly, the phase gradients maps $\partial_y\Phi$ and $\partial_x\Phi$ are calculated during the processing of the interferogram (from Figs. \ref{algoStructure}k,l), as an intermediary step before getting the OPD map \cite{JPDAP54_294002}. Consequently, $C$ and the noise level can be easily calculated in CGM, for any experimental image. Details of the mathematical procedure are given in Ref. \cite{OE21_17340}.}

\subsection{Effect of the illumination NA\label{sect:NA}}

So far, we considered a zero-NA illumination (plane wave, \NAi$=0$), also referred as a spatially \emph{coherent} illumination. When the NA of the illumination is increased, one usually states that spatial coherence of the light is decreased. The illumination NA has a notable effect on the OPD image in CGM. First, it enables a kind of sectioning in $z$ by blurring the out-of-plane parts of the imaged object \cite{OE22_8654}. Second, increasing \NAi, just like increasing $\mathsf{NA}_\mathrm{obj}$, leads to a better spatial resolution, a property advantageously used in Ref. \cite{BJ106_1588} .

However, increasing \NAi\ also tends to blur the interferogram, and thus to increase the noise amplitude of the image. Such an effect was investigated and explained by Bon et al. in Ref. \cite{OE17_13080} . More specifically, when a plane wave illuminates the sample with an incidence angle $\psi$, the plane wave transmitted through the optical microscope exhibits an incidence angle $\psi_\mathrm{t}$ at the image plane such that $\tan\psi_\mathrm{t}=\tan\psi/M$, where $M$ is the magnification of the microscope. This tilt angle $\psi$ results in a translation of the interferogram in one direction by a distance $d\tan\psi_\mathrm{t}=d\tan\psi/M$ (see the shadow picture in Ref. \cite{JPDAP54_294002}). When a non-zero illumination NA is used, several illumination angles $\psi$ enter into play and the final interferogram results from the incoherent sum of all the interferograms associated with all the illumination angles. Because they are all slightly shifted compared with each other on the camera plane, increasing the illumination NA results in a blurring of the interferogram, a lower contrast of the fringes and thus a higher noise on the OPD image. At some point, when \NAi\ reaches a critical values, the contrast is cancelled and the noise level diverges.\cite{OE17_13080}

The \emph{in silico} algorithm also offers the possibility to vary the illumination angle and, consequently, the illumination NA. A tilt angle can be easily applied during all the propagation steps (see Fig. \ref{algoStructure}) in the Fourier space. To model a given illumination NA, one just has to incoherently average all the interferograms corresponding to various illumination angles within the NA of the illumination, with a sufficient degree of angular discretisation.

\begin{figure*}
	\centering
		\includegraphics[scale=1]{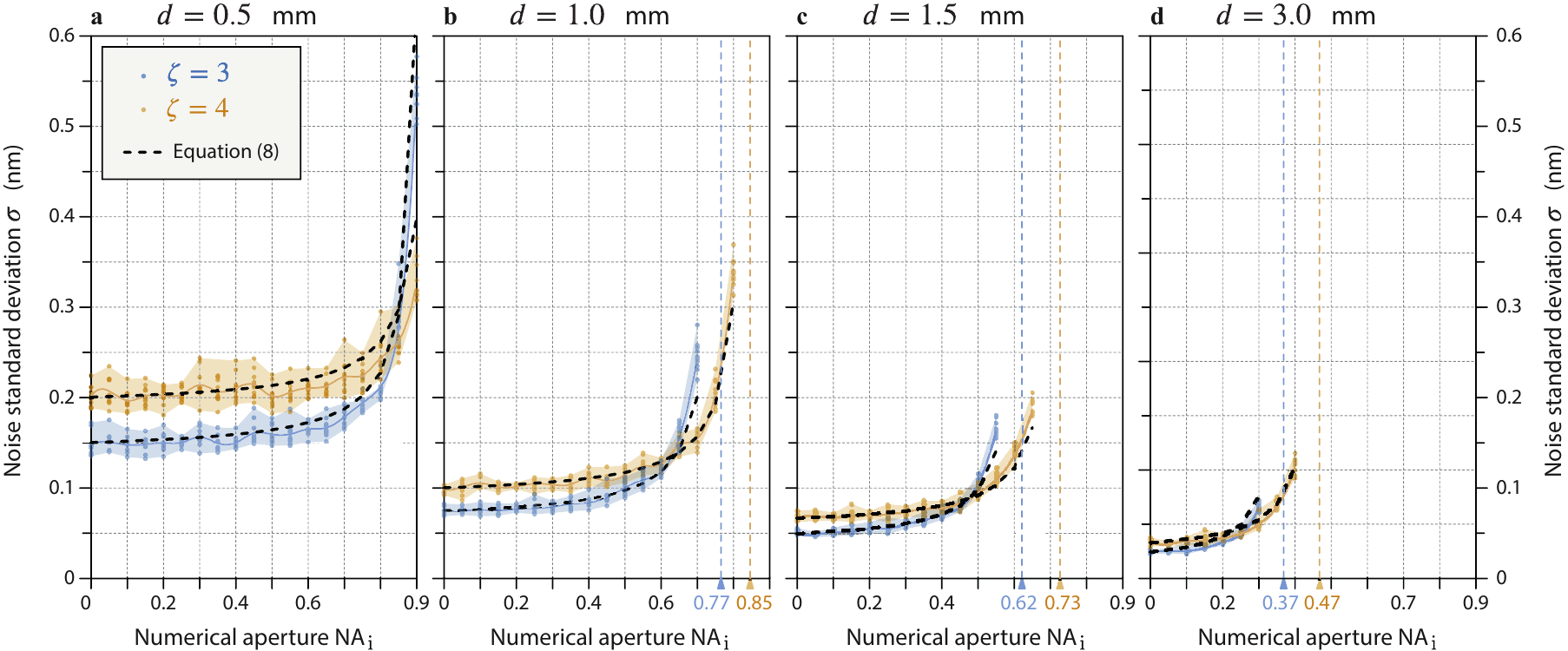}
	\caption{{\bfseries Effect of the illumination numerical aperture on the image noise.} Standard deviation of the noise as a function of the numerical aperture for grating-camera distances (a) $d=0.5$ mm, (b) $d=1.0$ mm, (c) $d=1.5$ mm, (d) $d=3.0$ mm. Each dot represents one \emph{in silico} measurement. Colored areas span from the minimum to the minimum values. Dashes lines represent $\sigma_0$ values calculated using Equation \eqref{eq:sigmaNA}. In (b,c,d), $\mathsf{NA}_0$ values (Eq. \eqref{eq:NA0}) are indicated by vertical dashed lines.}
\label{sigma_NA}
\end{figure*}

Figure \ref{sigma_NA} plots the noise standard deviation $\sigma$ of \insilex\ OPD images as a function of the illumination NA \NAi, for various camera-grating distances $d$. For small values, \NAi\ does not affect the image and Eq. \eqref{eq:sigma} giving the noise amplitude can be confidently used. However, for large NA, we observe the increase followed by a divergence of the noise amplitude, corresponding to a cancellation of the interferogram contrast. This cancellation and associated divergence occur for a very specific value of \NAi$=\sin\psi_\mathrm{i,max}$, as explained in Ref. \cite{OE17_13080}, such that
\begin{equation}
KR=1.22\pi
\label{KR1.22}
\end{equation}
where $K=4\pi/\Gamma$ and $R=d\tan\psi_\mathrm{i,max}$. Results of Fig. \ref{sigma_NA} confirm exactly this condition. We indeed observe divergences of $\sigma$ for \NAi\ values corresponding to Eq. \eqref{KR1.22} that we call the limiting NA $\mathsf{NA}_0$. In all the plots, the noise standard deviation $\sigma$ could be nicely fitted with a function of the form $f(\zeta/d)=a_\mathrm{fit}(\zeta/d)+b_\mathrm{fit}(\zeta/d)/(\mathsf{NA}_0(\zeta/d)-\mathsf{NA}_\mathrm{i})$. We found an expression of the limiting NA
\begin{equation}
\mathsf{NA}_0=\sin\left[\tan^{-1}\left(\frac{1.22M\Gamma}{4Zd}\right)\right]\label{eq:NA0}
\end{equation}
that exactly matches the condition \eqref{KR1.22} theoretically derived in Ref. \cite{OE17_13080} . Then, the values of $a_\mathrm{fit}$ and $b_\mathrm{fit}$ led us to the refined expression of $\sigma_0$ (refined Equation \eqref{eq:sigma}) involving the illumination NA:\vspace{2em}\\

\hfsetfillcolor{gray!8}
\hfsetbordercolor{black!50}
\begin{align}\tikzmarkin{a}(1.0,-0.6)(-0.9,0.75)
\hspace{-2em}\noindent\sigma_0=&\left(\frac{p\Gamma}{8\sqrt{2}Zd} + \frac{\mathsf{NA}_\mathrm{i}}{\mathsf{NA}_0-\mathsf{NA}_\mathrm{i}}\times3.54\; [\text{nm}] \right)\nonumber\\
&\times\sqrt{\left(\frac{1}{N_\mathrm{im}}+\frac{1}{N_\mathrm{im}^0}\right)\frac{log(N_xN_y)}{w}}
\tikzmarkend{a}\label{eq:sigmaNA}\end{align}
Figure \ref{sigma_NA} plots fits of all the \insilex\ data using Eq. \eqref{eq:sigmaNA}, showing an acceptable agreement.

\section{Trueness}
\label{sect:trueness}
While the measurement precision is easy to measure experimentally, the trueness cannot always be quantified. Estimating the trueness of a measurement requires the use of a calibrating sample. Interestingly, with the \insilex\ algorithm, the trueness of CGM as a function of all the experimental parameters can be easily estimated, because the true OPD image is known (Fig. \ref{algoStructure}f). 

Equation \eqref{eq:sigma} suggests that the image noise can be infinitely dampen upon infinitely increasing the grating-camera separation $d$. As expected, this law holds true up to a certain limit. The limit is the accuracy. If the grating is put far away from the sensor, then the noise is reduced but the range of wavefront gradients that can be quantitatively imaged is also reduced. If the wavefront gradient is too steep, the integration algorithm yields incorrect reconstruction. As a consequence, although the noise can be infinitely reduced, one cannot infinitely increase the signal to noise ratio in CGM in a safe manner just by playing with the grating-camera distance.

To quantify this limitation, we ran \insilex\ experiments on a model Gaussian OPD profile of amplitude $A$ for a large range of $d$ and $\Gamma$ values. For each set of $d$ and $\Gamma$ parameters, a loop in $A$ values (Figure \ref{algoStructure}f) was run until a discrepancy of 5\% was observed between the actual $A$ value and the one of the reconstructed OPD image (Figure \ref{algoStructure}m). This procedure enabled the phenomenological determination of the $(d,\Gamma)$ association of values that yield discrepancy. This set of values corresponds to a limiting wavefront gradient on the camera that reads:
\begin{equation}
|\nabla_\mathrm{c} W|_\mathrm{max}=\psi^\mathrm{c}_\mathrm{max}=\frac{\Gamma}{4Zd}=\frac{\Gamma'}{4d'}
\label{eq:nablaWmax}
\end{equation}
$\nabla_\mathrm{c}$ means the gradient at the image plane. Note that the gradient of the wavefront $\nabla_\mathrm{c} W(x,y)$ is nothing but the local angle of incidence of the wavefront on the image plane at $(x,y)$. Thus, equation \eqref{eq:nablaWmax} gives the expression of the maximum angle of incidence $\psi^\mathrm{c}_\mathrm{max}$ a wavefront can have on the camera to be properly characterized. This limited angle corresponds to a shift of a spot by half a spot size in the interferogram (the spot size being $\Gamma'/2$), i.e. the situation where a bright spot of the interferogram impinges on a neighboring spot location. Chanteloup et al. discussed this limitation in Ref. \cite{SPIE5252}. They, however, gave an expression of the limiting angle that differs by a factor of 4 from our result.

In practice, one rather deals with the wavefront profile at the sample plane, not at the image plane. At the sample plane, the OPD amplitude remains unchanged, however, the wavefront is laterally shrinked and its gradient is thus magnified, by a factor of $M$, the magnification of the microscope. This condition yields the incidence angle threshold (IAT) at the sample plane:
\begin{equation}
|\nabla \delta\ell|_\mathrm{max}=\psi_\mathrm{max}^\mathrm{s}=\frac{M\Gamma}{4Zd}\label{phimaxs}
\end{equation}

\begin{figure*}
\centering
\includegraphics[scale=1]{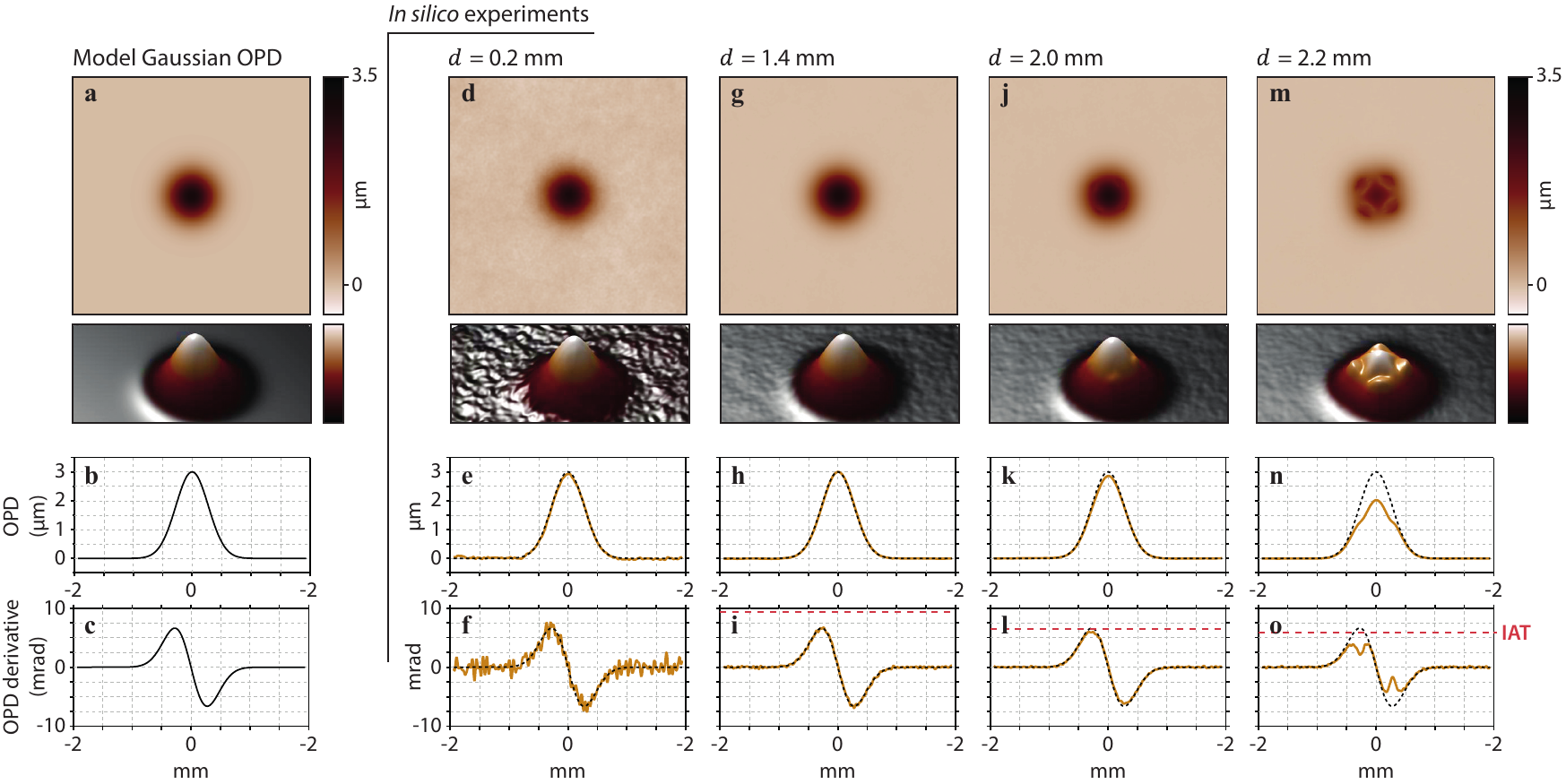}
\caption{{\bfseries Characterization of the accuracy.} (a) Model Gaussian OPD distribution at the camera plane ($3.9\times3.9$ mm$^2$ area), along with a 3D rendering of the image. (b) OPD cross-cut at the center of image (a). (c) Derivative of (b). (d,e,f) Same as (a,b,c) for an in silico experiment with a grating-camera distance of $d=0.2$ mm. (g,h,i) Same as previously for a  grating-camera distance of $d=1.4$ mm. (j,k,l) Same as previously for a  grating-camera distance of $d=2.0$ mm. (m,n,o) Same as previously for a  grating-camera distance of $d=2.2$ mm.  The dotted lines in (e,h,k,n) recall the model cross cut (b). The dotted lines in (f,i,l,o) recall the model derivative cross cut (c). The red dash lines in (i,l,o) indicate the incidence angle threshold (IAT) $\psi_\mathrm{max}$.}
\label{accuracy}
\end{figure*}

To better observe the nature of the problem that arises when the limit is reached, we conducted simulations at a specific $\Gamma$ value, and vary the distance $d$. Results are presented in Figure \ref{accuracy}. The model object was a Gaussian OPD distribution (Figure \ref{accuracy}a), impinging on the camera, $A=3$ \textmu m in amplitude and $0.65$ mm in full-width half-maximum (at the image plane, so that the IAT is given by Eq. \eqref{eq:nablaWmax}). The OPD distribution is passed through the \insilex\ algorithm to compute experimental images for different grating positions $d=0.2, 1.4, 2.0, 2.2$ mm. The maximum counts on the camera has been set to a low value of $w=200$ to highlight the effect of $d$ on the noise. For $d=0.2$ mm, a perfect agreement is found between model and \insilex\ experiment, but a high noise level is also observed (Figs. \ref{accuracy}d,e,f). In agreement with Eq. \eqref{eq:sigma}, this noise is reduced upon increasing $d$ (Figs. \ref{accuracy}g-o). However, at $d=2.0$ mm, a small discrepancy is observed in the \insilex\ Gaussian profile (Fig. \ref{accuracy}k), an inaccuracy that becomes dramatic at $d=2.2$ mm (Fig. \ref{accuracy}n). This last example corresponds to the case where the wavefront profile locally exceeds the IAT at some places in the image. Figures \ref{accuracy}l,o plot the incidence angle (i.e., the derivative of the OPD at the image plane) for the two problematic cases and we can see that the problem arises when the wavefront derivative reaches the IAT (dashed, red lines). When such an issue occurs experimentally, it usually leads to 4-fold symmetry artefacts (Fig. \ref{accuracy}m), aligned with the grating tilt angle $\theta$.  Such an issue typically arises upon imaging objects with sharp boundaries, or with nanoparticles. The effect is even stronger when the surrounding medium has a low refractive index (typically nano- or micro-beads in air). We personally encountered this issue when imaging micro-bubbles in liquids, and, under some conditions, 1-\textmu m dielectric beads in air and gold nanostructures. To lift this problem, according to equation \eqref{phimaxs}, either the grating has to be put closer to the camera, or the microscope magnification $M$ has to be increased to expand the wavefront on the camera and diminish its gradient $\psi^\mathrm{c}$. {\revision  Also, the numerical aperture in detection, $\mathsf{NA}_\mathrm{obj}$, can be reduced to smooth the image and soften the gradients.}\\

{\revision The quality of the OPD image reconstruction in CGM was also studied recently by Stolidi et al., in the X-ray spectral range \cite{OE30_4302}. The authors introduced a so-called confidence map, stemming from the PDCM $C$ defined by Eq. \eqref{eq:PDCM}, which can tell which parts of the reconstructed OPD image suffer from artefacts coming from undersampling, like in Fig. \ref{accuracy}m. This approach amounts to highlighting OPD derivatives that are close to the IAT we define in this article.}

{\revision 
\section{CGM parameters optimization}
The important question we shall now address in this section is how to set all the parameters of a home-made CGM system to achieve the most accurate measurements. To answer this question, both Eqs. \eqref{eq:sigma2} and \eqref{phimaxs} must be considered.

First, Eq. \eqref{eq:sigma2} involves three sets of parameters, those related to the cross-grating ($\zeta$, $d$), the one related to the relay lens ($Z$), and those related to the camera ($p$, $N_i$, $w$). Hence, this equation is important as it represents a guide to select or design the optimal camera, relay-lens and cross-grating to achieve the lowest noise amplitude.

Regarding the cross-grating, $\zeta$ has to be as small as possible. The smallest possible value begin $2.73$, to satisfy Nyquist criterion, as explained above. In practice, it ranges from 3 to 4. Regarding the grating-camera distance $d$, the further the better but there exists a tradeoff as explained in Sect. \ref{sect:trueness}. The distance must remain below a given value that depends on the imaged object to avoid inaccuracies. If we consider that the distance is optimized to reach the IAT: $d=M\Gamma/4Z\psi_\mathrm{max}^\mathrm{s}$, then Eq. \eqref{eq:sigma2} reads
\begin{equation}
\sigma_0=\frac{p \psi_\mathrm{max}^\mathrm{s}}{2M}\sqrt{\frac{\log(N_xN_y)}{N_\mathrm{im}w}}\label{eq:sigma3}
\end{equation}

With this new expression of $\sigma_0$, the grating and relay-lens parameters fully disappear, and we are left only with the two important camera parameters: the pixel size $p$ and the full well capacity $w$. More precisely, the only physical quantity that matters, according to Eq. \eqref{eq:sigma3}, is the ratio $p/\sqrt{w}$ that should be as small as possible. In other words, the ratio
\begin{equation}
\rho=\frac{w}{p^2}
\end{equation}
must be chosen as large as possible. Interestingly, this ratio $\rho$ is the maximum areal charge density that the camera sensor can hold. Naturally, this number is quite constant from one camera to another, because the quantity of photoelectron $w$ a camera dexel can contain is roughly proportional to its area $p^2$. This value lies in the range of 300 to 1200 e$^-/$\textmu m$^2$ for common scientific cameras. As a consequence, the choice of the camera is not of primary importance to achieve precise measurements. Any low cost camera should already yield acceptable measurements in CGM, provided the highest dynamic range of the camera is selected (12-bit, 16-bit, ...) to actually benefit from the full well capacity.

A camera feature that is indirectly involved in Eqs. \eqref{eq:sigma}, \eqref{eq:sigma2} and \eqref{eq:sigma3} is the frame rate $f$. Indeed, large frame rates enable the acquisition of more images per unit of time, i.e., higher values of $N_\mathrm{im}$. High-speed camera may be a means to achieve better signal to noise ratio, keeping in mind that higher frame rate comes along with shorter exposure time and the use of brighter light sources to keep on benefiting from the full well capacity of the camera sensor. Using high illumination intensity may be detrimental when studying living cells.

Finally, Eqs. \eqref{phimaxs} and \eqref{eq:sigma3} highlights the importance of the objective magnification. Using high magnification objectives enables one to position the grating further from the image plane, and to reduce the noise amplitude. This is why it appears in the denominator of Eq. \eqref{eq:sigma3}.}

\section{Summary and perspective}
In this article, we provide a numerical procedure to simulate experimental measurements in cross-grating phase microscopy (CGM). With such \emph{in silico} experiments, it becomes possible to vary all the parameters involved in CGM experiments, including those that can hardly be varied in practice (or even cannot, with built-in systems), such as the camera-grating distance, grating pattern, camera pixel size, grating period, zoom of the relay lens, grating angle, etc. Importantly, the algorithm takes into account the noise formation of the reconstructed OPD images, the main concern to conduct challenging CGM experiments, at the limit of the state of the art, for instance for application in nanophotonics, or in microbiology where the objects of interest can be very thin, small and below the diffraction limit. The algorithm can help saving a considerable amount time and data storage space, by varying many parameters within numerical loops, instead of running actual experiments for months, in order to guide the adjustment of a CGM setup and the effective realization of actual experiments.

To illustrate the interest of the algorithm, we applied it to deepen the understanding of noise formation in CGM, and its implication in terms of precision and trueness. In particular, we derived fundamental expressions of the noise standard deviation and of the wavefront gradient limit, as a function of all the CGM parameters. However, the possibilities of this algorithm go much beyond. For instance, few investigations have been made in the improvement of the grating pattern, which has remained identical for 20 years (QLSI grating), especially because it would imply the expensive fabrication of a large amount of cross-gratings, and the associated masks for photolithography. We focused here on this popular QLSI pattern, but with the \insilex\ algorithm, one just has to modify a matrix to investigate and quantify the interest of any new grating pattern. Also, because actual noise is reproduced, the algorithm could be a priori used to easily build an arbitrarily large library of images as the ground truth to train segmentation or denoising deep-learning algorithms.

\section{Acknowledgement}
This project received funding from the European Research Council (ERC) under the European Union’s Horizon 2020 Research and Innovation Programme (grant agreement no. 772725, project HiPhore). The authors acknowledge support for the SATT Sud-Est.

\section{Bibliography}

\bibliographystyle{unsrtnat}


\end{document}